\documentclass[12pt]{iopart}

\usepackage{graphicx}

\begin{document}

\title{Five-Fold Reduction of Lasing Threshold near the First $\Gamma L$-Pseudogap of ZnO Inverse Opals}

\author{Michael Scharrer$^{1}$, Heeso Noh$^{1,2}$\footnote{corresponding author}, Xiaohua Wu$^{1}$, Mark A Anderson$^{1}$, Alexey Yamilov$^{3}$, Hui Cao$^{1,2}$ and Robert P H Chang$^{1}$}
\address{$^1$Materials Research Institute, Northwestern University, Evanston, IL 60208}
\address{$^2$Dept. of Applied Physics, Yale University, New Haven, CT 06511}
\address{$^3$Dept. of Physics, Missouri University of Science and Technology, Rolla, MO 65409  }

\begin{abstract}

We report room temperature lasing in ZnO inverse opal photonic crystals in the near-ultraviolet (UV) frequency. We observe random lasing due to disorder in the structures when the photonic pseudogaps are located away from the ZnO gain spectrum. Tuning the first $\Gamma L$-pseudogap to the gain peak leads to a five-fold reduction in lasing threshold and frequency shift of lasing modes due to the enhanced confinement of light.

\end{abstract}


The pursuit of light localization has led to two alternative approaches to realize mirrorless lasers: random lasers and photonic crystal (PhC) lasers.  In a random laser, feedback is provided by strong scattering of light in a disordered medium.  Random lasing has been observed in various disordered media.\cite{cao_lasing_2003,noginov_solid_2005,cao_review_2005, wiersma_physics_2008}  However, the currently achievable thresholds are too high for many practical applications because of incomplete confinement of light.  One approach to improve the confinement is to maximize the scattering strength by using Mie resonances.\cite{wu_random_2004,Gottardo_2008}  Another approach, first proposed by John,\cite{john_strong_1987} is to reduce the effective momentum of light by introducing periodicity into the system: The Ioffe-Regel criterion for light localization in the presence of periodicity is replaced by $k_c\ell\leq 1$, where $k_c$ is crystal momentum, which is much smaller than the optical wavevector $k$ at near band edge. Due to the challenges in fabrication, real PhCs possess an unavoidable degree of disorder, and the optical properties of such partially ordered systems must be understood because uncontrolled scattering is very detrimental to passive PhC devices.\cite{vlasov_different_1999,hughes_2005,koenderink_optical_2005} Theoretical studies of 2D systems have shown that very high quality modes can exist in disordered PhCs \cite{yamilov_highest-quality_2004,yamilov_self-optimization_2006, rodriguez_disorder-immune_2005}, which facilitates lasing action. A recent study illustrates a gradual transition from lasing in a photonic crystal to random lasing behavior.\cite{kwan_transition_2007}  In this chapter we report experimental results on the UV lasing characteristics of ZnO inverse opal PhCs. We observe a strong reduction in lasing threshold when the fundamental PBG in the [111] direction is tuned to overlap with the gain spectrum of ZnO.  This demonstrates a combination of random lasing with partial PBG confinement.

In 3D PhCs, realizations of gain enhancement,\cite{vlasov_enhancement_1997,maskaly_amplified_2006} stimulated emission\cite{romanov_stimulated_2004}, and lasing\cite{yoshino_amplified_1999,cao_lasing_2002,shkunov_tunable_2002}  so far have relied on light sources infiltrated into a passive PhC.  Infiltration complicates fabrication and can lead to a reduction of refractive index contrast and interactions of the emitters with the dielectric walls.  In our samples ZnO acts as both the dielectric backbone and the gain medium for lasing.  This allows us to study the emission properties without infiltrating quantum dots or dye molecules.  However, we need to take into account the frequency-dependence of the refractive index and absorption in the PBG region in our active systems. 

\section{Fabrication of ZnO inverse Opal}

\begin{figure}[htbp]
\center
\begin{tabular}{c}
\includegraphics[width=3in]{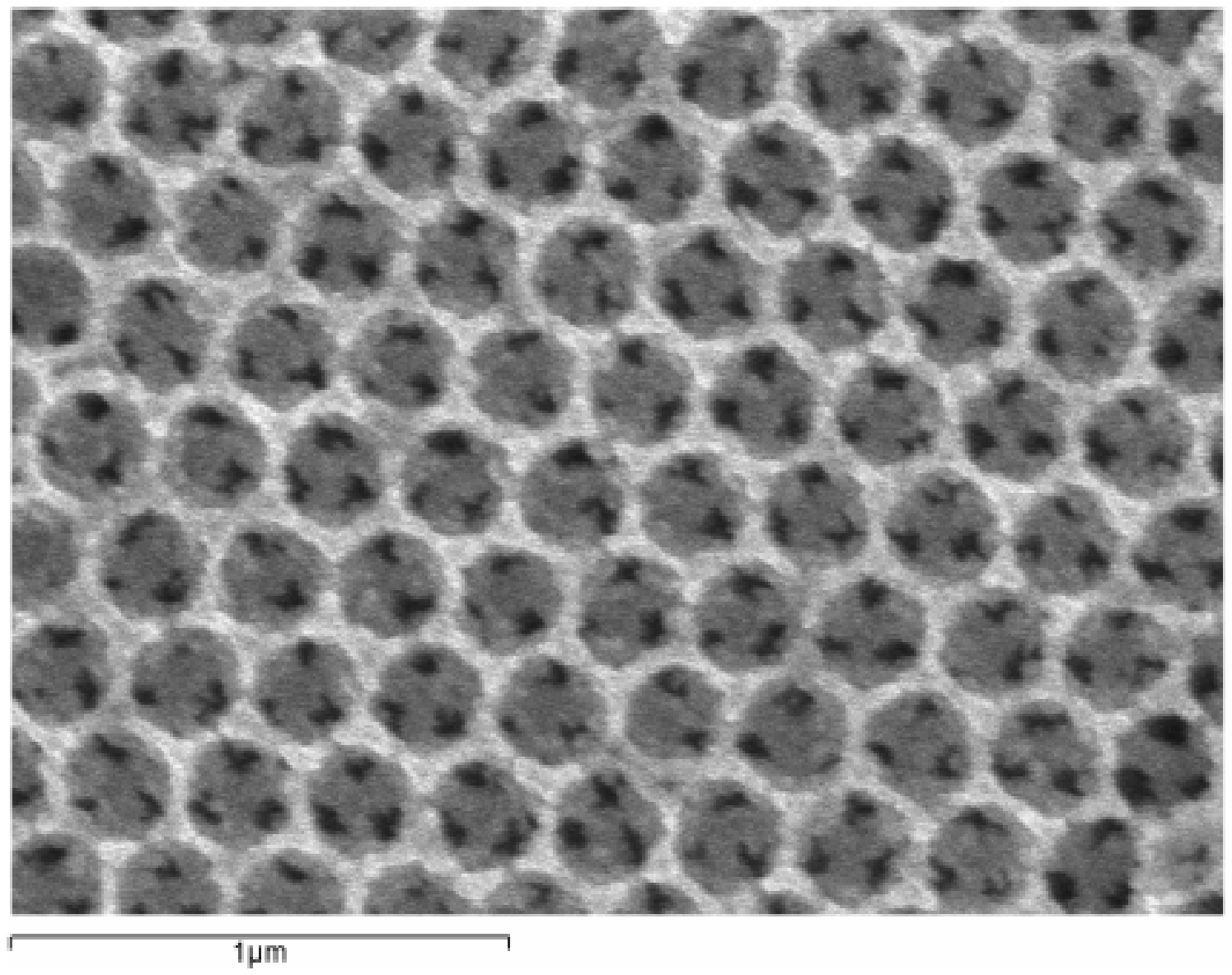}\\
(a)\\
\includegraphics[width=3in]{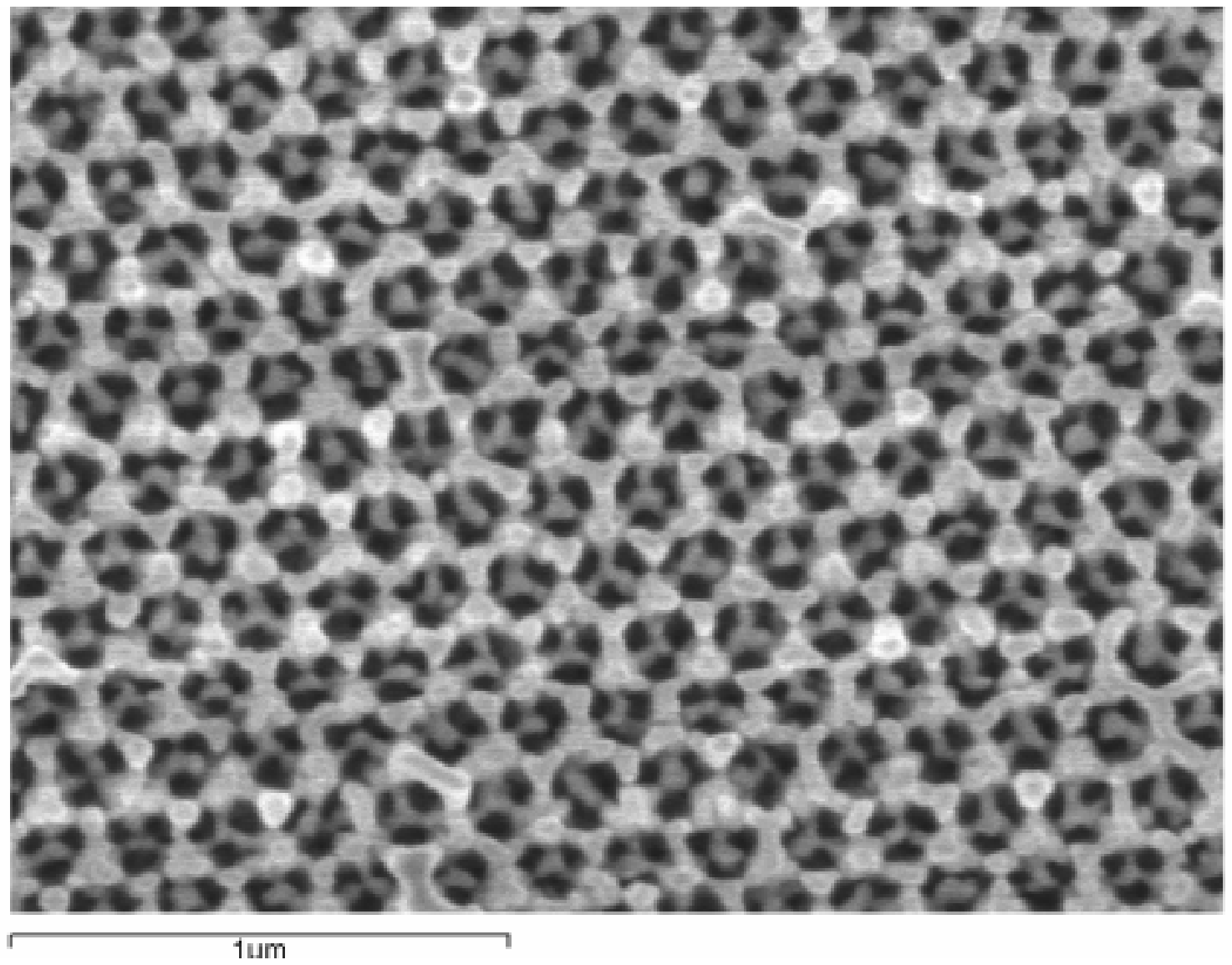}\\
(b)\\
\end{tabular}
\caption[SEM images of ZnO inverse opals]{ Scanning electron microscope (SEM) images of ZnO inverse opals. (a) (111) surface of d=$256nm$ sample and (b) cross section (cleaved surface) of d=$171nm$ sample.  The rough granular structure is due to crystal growth during firing. }
\label{fig:UV-SEM}
\end{figure}

Polystyrene opals were prepared by self-assembly onto glass substrates using the vertical deposition technique. The opals are typically $\sim 50$ layers thick with single crystal domains of several tens by hundreds of micrometers.  The templates were infiltrated with ZnO by atomic layer deposition (ALD) and then removed by firing at $550^\circ C$.  The remaining structures are face-centered cubic (FCC) arrays of air spheres surrounded by ZnO dielectric shells, with the (111) crystal surface parallel to the substrate [figure \ref{fig:UV-SEM} (a)].  All samples were infiltrated and fired together to ensure similar material and optical qualities.

High-quality ZnO inverse opals have been reported with the first PBG in the red and  near-IR spectrum.\cite{jurez_zno_2005,scharrer_ultraviolet_2006} However, the small sphere diameters ($< 200nm$) necessary to tune the fundamental gap to the UV spectrum are difficult to grow with good monodispersity and to assemble into defect-free structures.  In addition, after infiltration firing at elevated temperatures causes sintering and grain growth of the nanocrystalline ZnO and thereby leads to small scale disorder in the structure due to roughness of the ZnO shell surfaces (figure \ref{fig:UV-SEM} (b)).  As a result, disorder becomes increasingly important with decreasing sphere size.

\break

\section{Optical Setup}

Reflection spectra of the PhCs were measured with a Cary 500 UV-vis-NIR spectrophotometer. Both the diffuse and the  total (diffuse plus specular) reflection from the sample surface are measured using an integrating sphere. The specular component can then be obtained by subtracting the two measured spectra. The angle of incidence of the beam onto the sample surface is $3^\circ$, i.e. very close to the normal [111] direction.  

\begin{figure}[htbp]
\center
\begin{tabular}{c}
\includegraphics[width=4in]{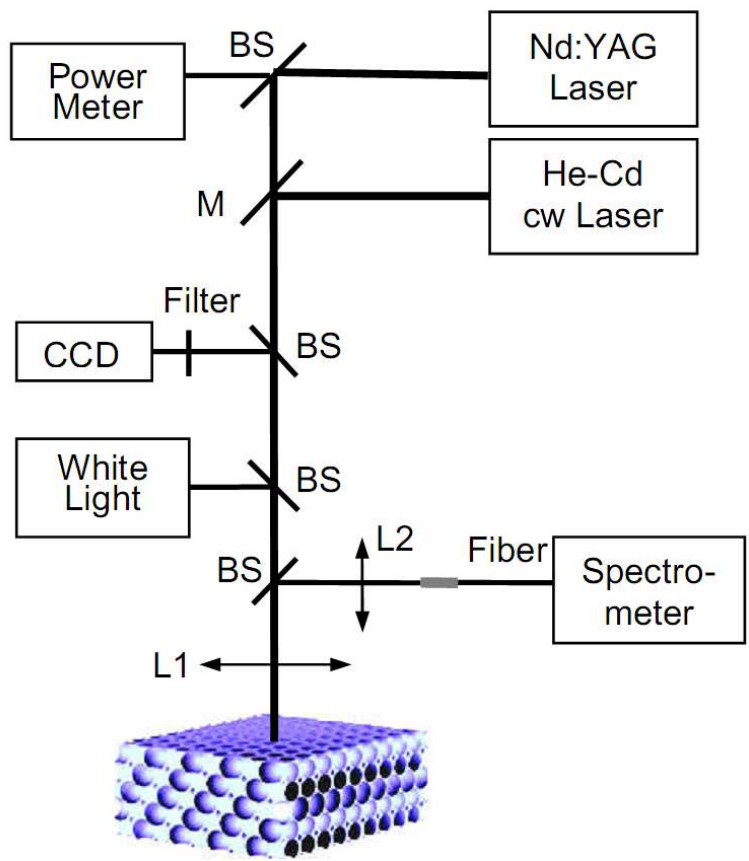}
\end{tabular}
\caption[Schematic of the optical setup used to measure lasing and PL]{ Schematic of the optical setup used to measure lasing and photoluminescence from the ZnO inverse opal structures.  The white light and pump beam are incident normal to the sample surface along the [111] direction, and the emission is collected in the same direction. BS stands for UV beam splitter, L1 is a 20X UV objective lens, M is a flip mirror to select the pump beams. }
\label{fig:chap2-setup}
\end{figure}

The optical setup used to measure lasing and photoluminescence is shown in figure \ref{fig:chap2-setup}.  To measure the emission from the samples, the sample surfaces were first imaged by a white-light  source  and  a  20X  objective  lens  onto  a  CCD  camera,  and  highly  reflective  areas  free  of cracks were selected for our experiments.  The samples were then pumped by a continuous-wave He-Cd laser of $\lambda=325nm$ in photoluminescence (PL) measurements, or at 10 Hz with 20ps pulses of wavelength $\lambda=355nm$ from a mode-locked Nd:YAG laser in the lasing experiment.  The beam had a spot diameter of approximately $20\mu m$ and was incident along the [111] crystal direction. The emission was collected by the objective lens (NA=0.4).  All measurements were performed at room temperature. 

\begin{figure}[htbp]
\center
\includegraphics[width=4in]{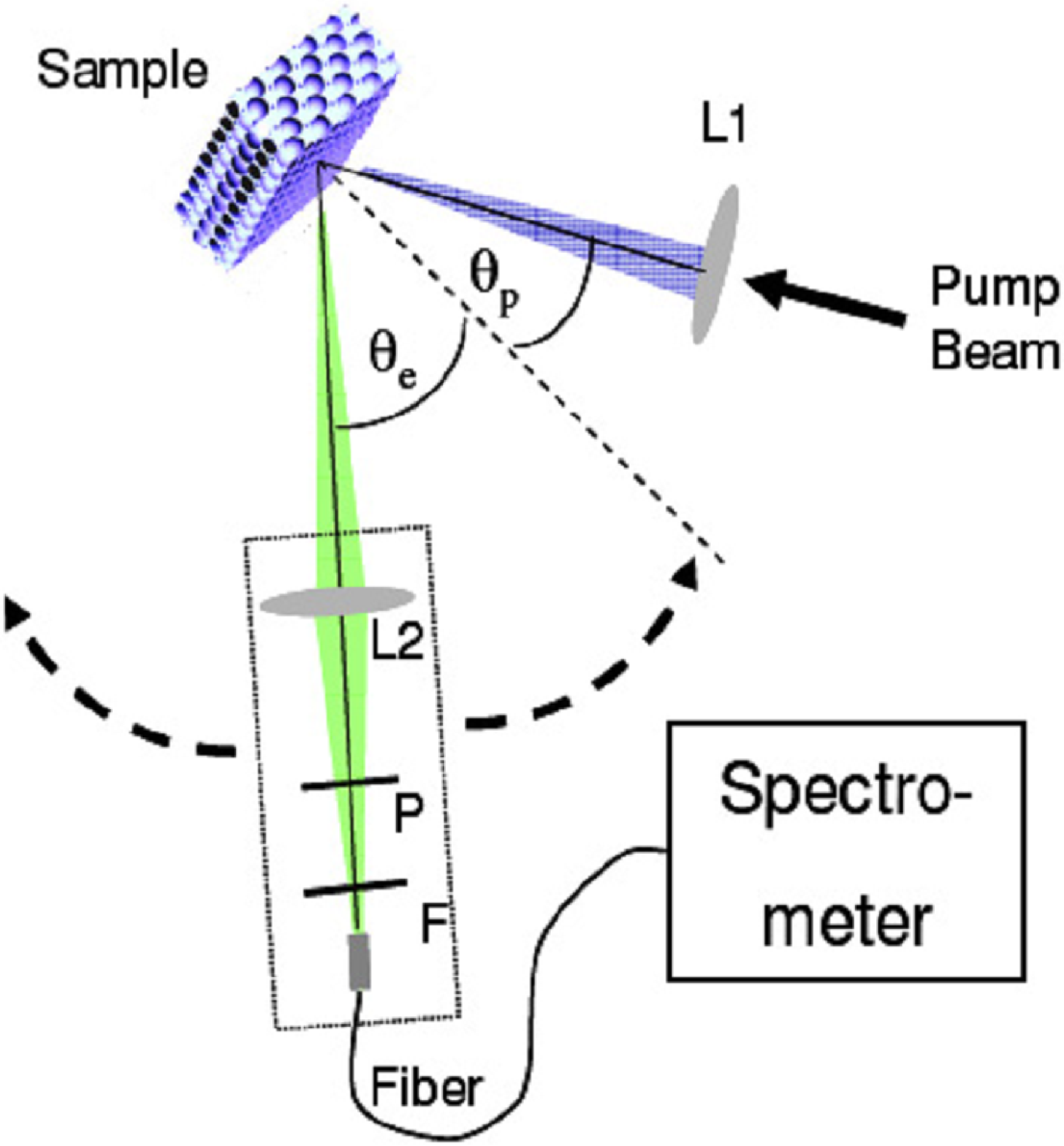}
\caption[Schematic of the optical setup used to measure angle resolved lasing]{Schematic of the optical setup used to measure angle resolved laser emission. The sample was mounted on a goniometer stage. Only the detection arm moved and the sample did not rotate. The pump beam was focused onto the sample by lens (L1). Emission was collected by another lens (L2) and focused to optical fiber bundle connected to a spectrometer. Polarization was selected by linear polarizer (P). A long pass filter (F) was placed in front of the fiber bundle to block the scattered pump light. }
\label{fig:setup_1}
\end{figure}

The angle resolved lasing experiment setup is shown in figure \ref{fig:setup_1}, The sample was mounted on a goniometer stage. Only the detection arm moved and the sample did not rotate. The third harmonics of the pulsed Nd:YAG laser was used to pump the ZnO. The pump beam was focused onto the sample focused by a
lens (L1) at a fixed angle $\theta_p \sim 30^{\circ}$. The emission was collected by another lens (L2) and focused to a fiber bundle which was connected to a spectrometer. A linear polarizer (P) was placed in front of the fiber
bundle to select $s$- or $p$-polarized light with electric field perpendicular or parallel to the detection plane (consisting of the detection arm and the normal of sample surface). The angular resolution, which was determined mainly by the range of collection angle of lens L2, was about $5^{\circ}$. Spectra of laser emission into different angles $\theta_e$ were measured when the detection arm was scanned in the horizontal plane. To prevent the reflected pump light from entering the detector, the incident beam was slightly tilted vertically so that the incidence plane deviated from the detection plane. A long pass filter (F) was placed in front of the fiber bundle to block the scattered pump light at wavelength $\lambda = 355 nm$.

\break

\section{Results and Discussions}

\begin{figure}[htbp]
\center
\begin{tabular}{c}
\includegraphics[width=4in]{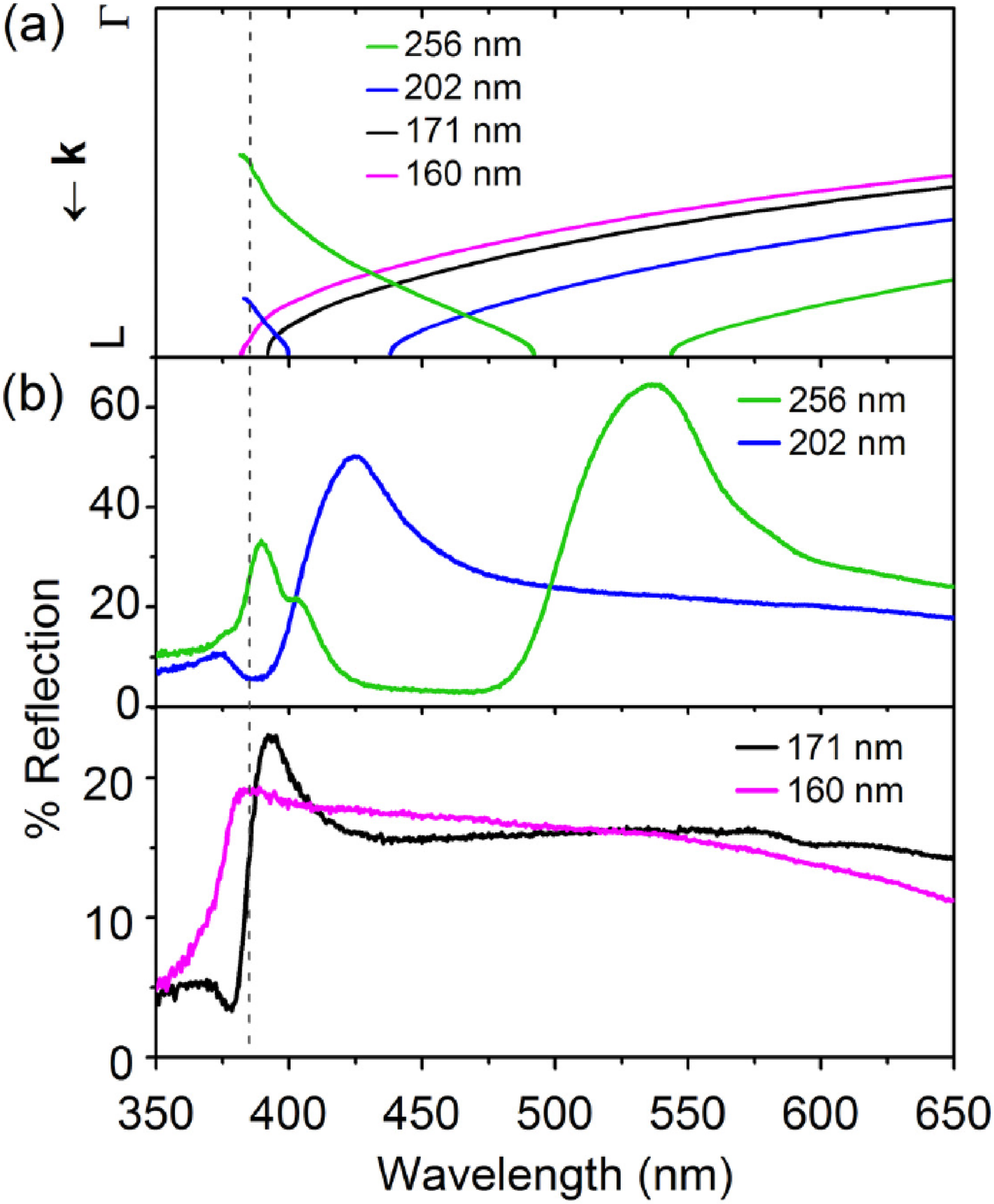}
\end{tabular}
\caption[Calculated photonic band structures and specular reflection spectra]{ (a) Calculated photonic band structures of the ZnO inverse opals in the vicinity of the first $\Gamma$L pseudogap.  The dashed line marks the approximate position of the ZnO absorption edge.  With decreasing sphere size the fundamental PBG shifts closer to the absorption edge and the $160nm$ sample has no PBG due to absorption.  (b) Specular reflection spectra of ZnO inverse opal PhCs with varying sphere diameters.}
\label{fig:chap2-gap}
\end{figure}

 Figure \ref{fig:chap2-gap} shows specular reflection spectra taken from four samples, with white light incident normal to the sample surface along the [111] crystal direction, and the calculated band structures of the PhCs. The largest sample ($d=256nm$) shows a main reflection peak at $\lambda=525nm$. It corresponds to the fundamental PBG in the (1 1 1) direction. Its wavelength is far from the ZnO absorption/emission edge. Additional reflection features
between $390$ and $400nm$ correspond to high order gaps. The narrow width and small amplitude of the reflection peak of the $171nm$ sample suggest that the fundamental gap is reduced by disorder and overlaps with the absorption edge of ZnO.  This means that the observed reflection peak corresponds to the low-frequency part of the PBG and the high-frequency part of the gap is effectively destroyed due to absorption by ZnO.  For the smallest sample
($d=160nm$) the fundamental PBG lies in the absorption region of ZnO and no PBG can be observed.
The calculations were performed using the plane-wave-expansion (PWE) method with values for the refractive index n($\lambda$) of ZnO for wavelength $\lambda>385nm$.\cite{scharrer_ultraviolet_2006} The calculated gap position
agrees well with the reflection peak frequency. 

\begin{figure}[htbp]
\center
\begin{tabular}{c}
\includegraphics[width=4in]{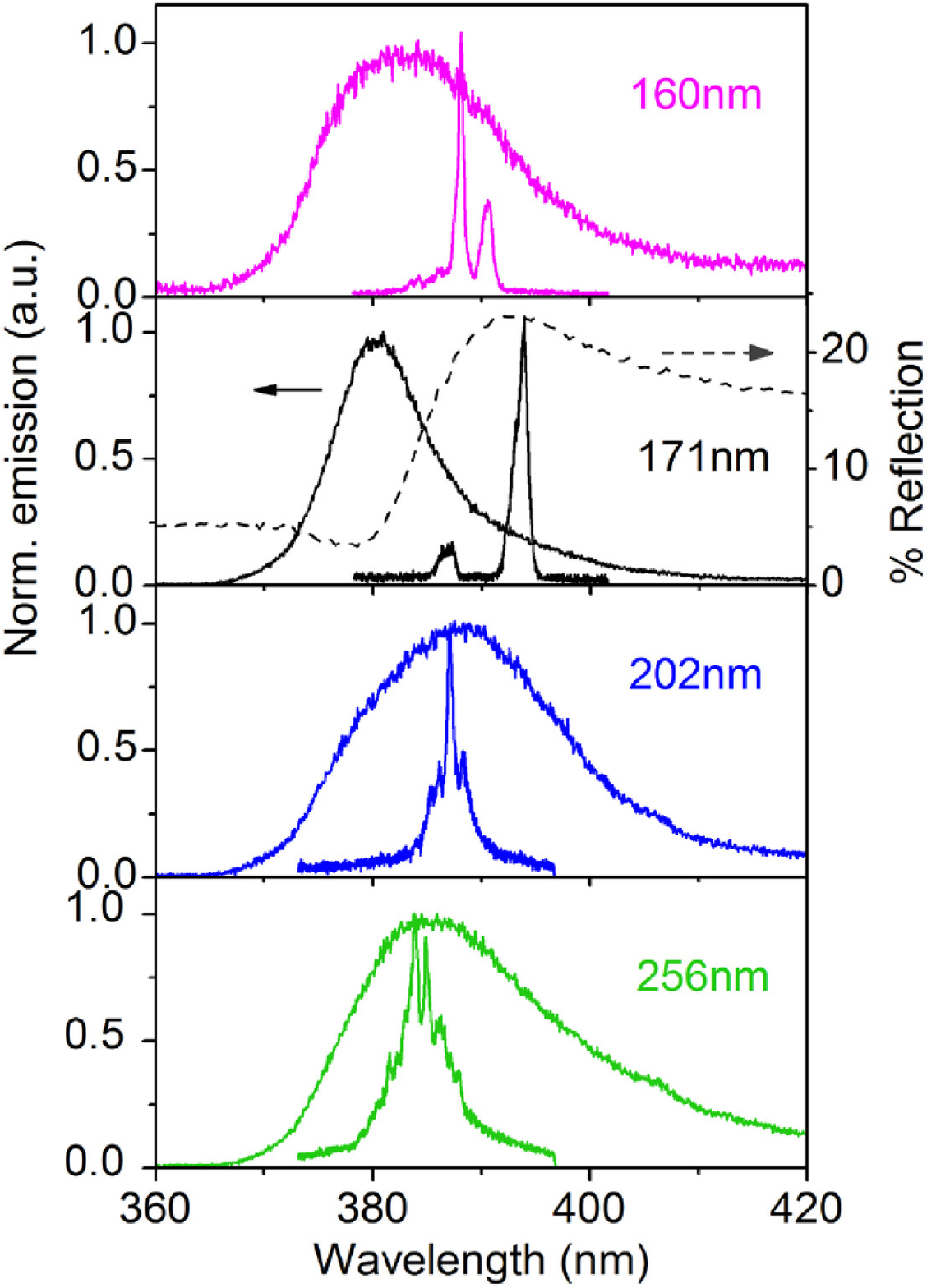}
\end{tabular}
\caption[Photoluminescence and lasing spectra]{Photoluminescence and lasing spectra of the ZnO inverse opals with varying sphere diameters.  The random lasing modes in the $d=160nm$, $202nm$ and $256nm$ samples overlap with the peak in PL spectrum.  In the $d=171nm$ sample the PL is suppressed and blue-shifted by the PBG (indicated by the reflection spectrum, dashed line). The main lasing modes are red-shifted into the PBG and do not overlap with the PL maximum. }
\label{fig:chap2-emission}
\end{figure}

The results of photoluminescence measurements are shown in figure \ref{fig:chap2-emission}.  We observe broad spontaneous emission peaks from the ZnO PhCs.  For a comparison of spectral shapes and peak positions, the emission
spectra have been normalized to a peak value equal to 1.  For the $d=171nm$ sample the PBG overlaps the emission band and the PL peak is clearly suppressed at the low-frequency edge and the maximum is blue-shifted. A similar
modification of the spontaneous emission near the first PBG has been observed for light-sources infiltrated into PhC structures and can be explained by a redistribution of emission from the directions prohibited by the PBG to other
allowed directions.\cite{koenderink_experimental_2003}

With increasing pump intensity, scattering of light by disorder in the structure causes random lasing in samples with no PBG ($d=160nm$) or a PBG away from the gain spectrum ($d=256nm$) (figure \ref{fig:chap2-emission}).  Similar random lasing behavior has been observed in ZnO powders.\cite{wu_random_2004,cao_random_1999} The random lasing modes overlap spectrally with the PL peaks and have output in many directions.  The linewidth of individual lasing peaks is $\sim 0.5nm$, but often several lasing peaks overlap partially in the spectrum.

\begin{figure}[htbp]
\center
\begin{tabular}{c}
\includegraphics[width=3in]{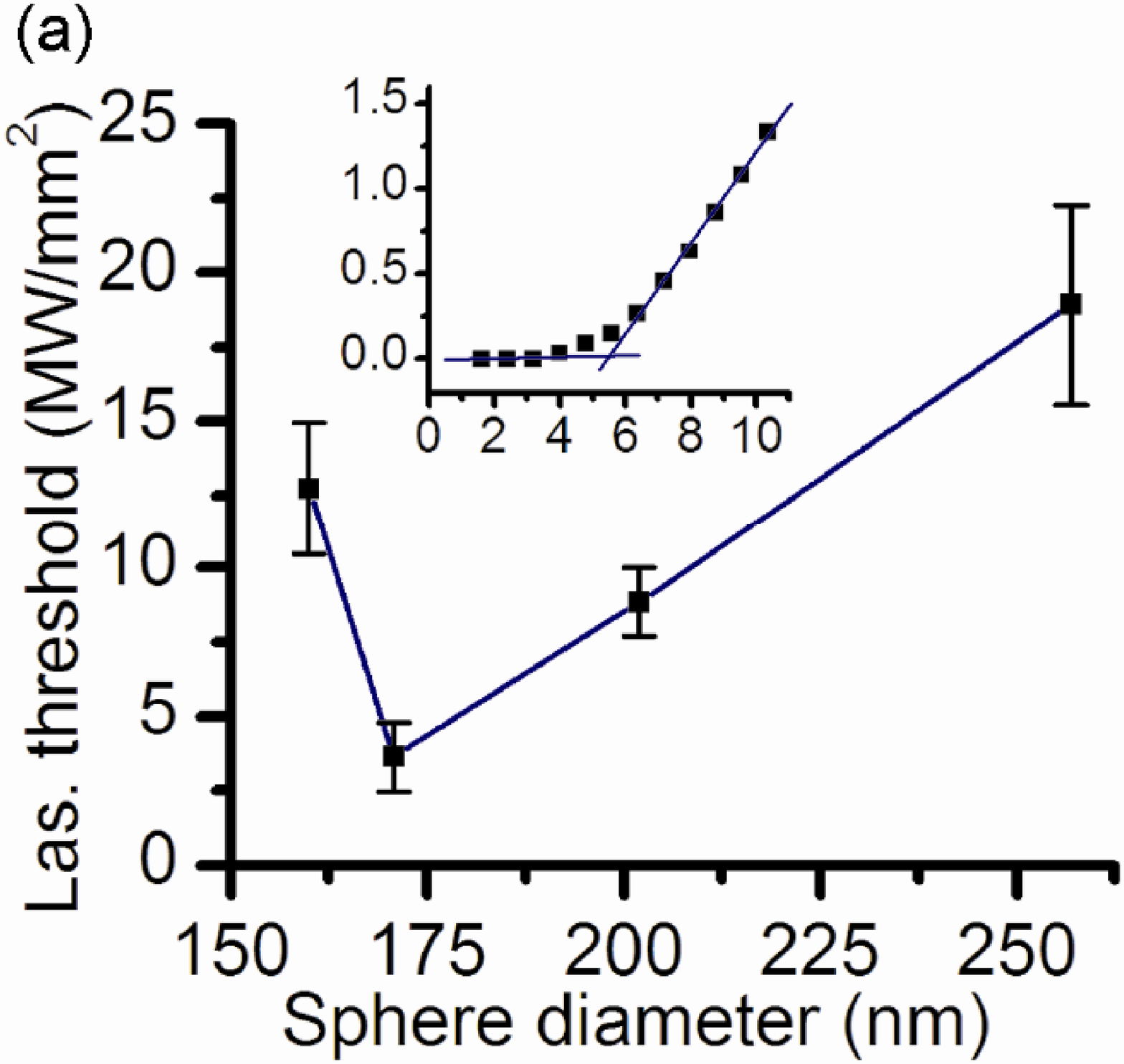}\\
\includegraphics[width=3in]{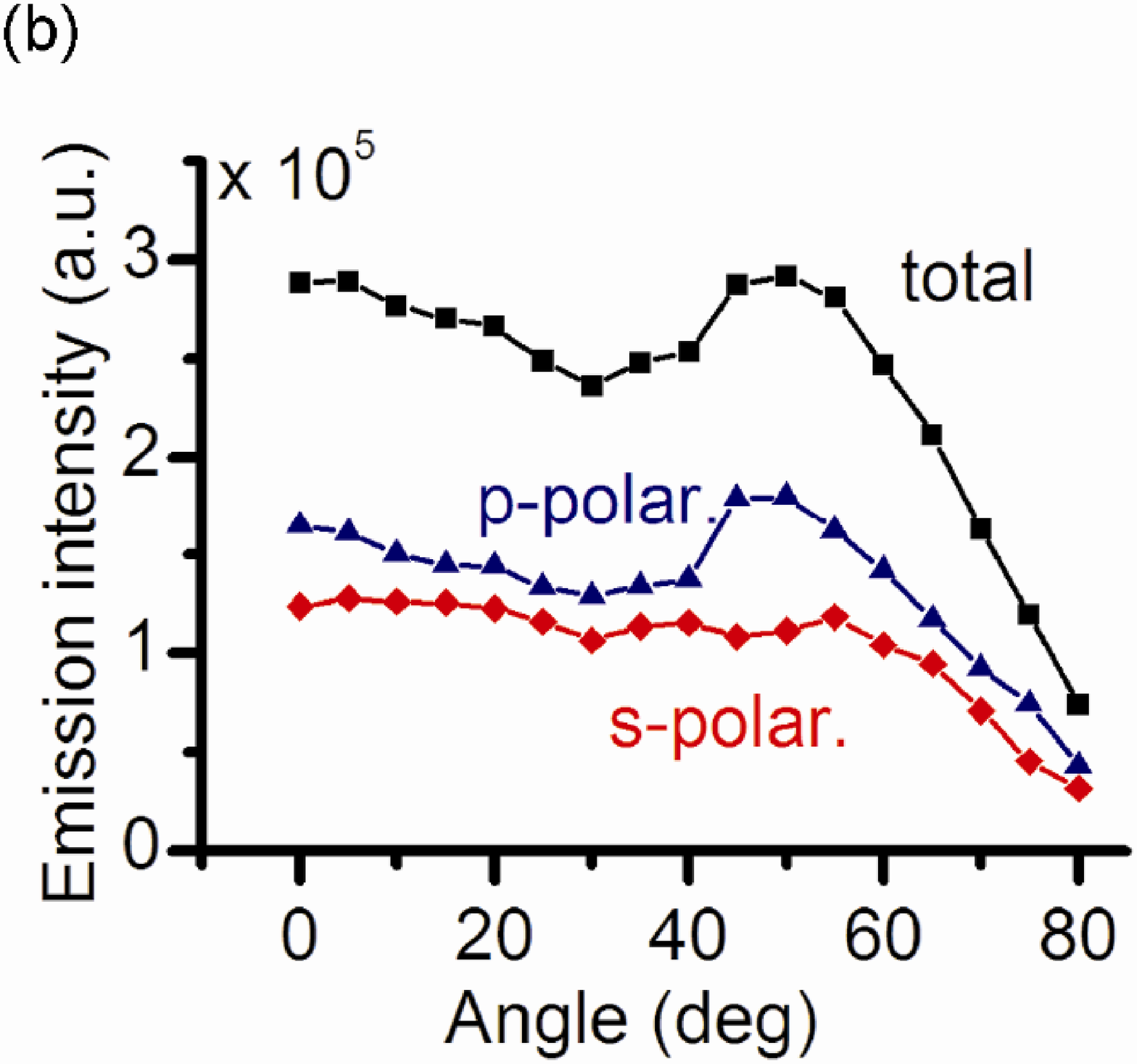}\\
\end{tabular}
\caption[Lasing threshold versus sphere diameter]{Lasing threshold versus sphere diameter for the ZnO inverse opals. The lasing threshold is strongly reduced when the PBG overlaps the emission spectrum of ZnO.  Inset: L-L curve measured for a $d=171nm$ sample, with lasing intensity (in arbitrary unit) plotted versus.pump intensity (in $MW/mm^2$).  (b) Lasing output intensity versus detection angle for the $d=171nm$ sample.  Lasing is not confined to the [111] direction.  The peak in p-polarized emission around $50^\circ$ is attributed to the Brewster angle effect.}
\label{fig:lasing}
\end{figure}

We have estimated the scattering mean free path of light $\ell_s$ in our samples from transmission and reflection data:  

\begin{equation}
\frac{T}{(1-R) \cdot T_{sg}\cdot T_{ga}}=exp(-\frac{L}{\ell_s})
\end{equation}

The probe light is incident onto the sample/air interface, T and R are the measured values of ballistic transmission and total reflection (including specular and diffusive reflection), $T_{sg}$ and $T_{ga}$ are the transmission
coefficients for the sample/glass and glass/air interfaces, and L is the sample thickness.  A comparison of $\ell_s$ values  for different samples is only possible at wavelengths away from the ZnO absorption band and the PBGs. Thus we
choose $\lambda=650nm$, and obtained values of $\ell_s$ equal to $29.2\mu m$ for $d=256nm$, $21.5\mu m$ for $d=202nm$, $22.8\mu m$ for $d=171nm$, and $23.3 \mu m$ for $d=160nm$. These results confirm that disorder slightly increases with decreasing lattice parameters, but also that the degree of disorder is similar for the three smaller sphere samples. Scattering will be significantly stronger at the ZnO emission wavelength due to increased refractive index of ZnO, but should remain comparable in all samples. 
  
Unlike the other samples, lasing peaks from the $d=171nm$ sample are spectrally located not at the position of strongest PL but are red-shifted into the PBG. As shown in figure \ref{fig:lasing}, lasing exhibits a clear threshold and the threshold decreases dramatically when the PBG is tuned to the gain spectrum of ZnO. More specifically, the threshold is reduced by approximately a factor of 5 from $d=256nm$ to $d=171nm$ and a factor of 3 from $d=160nm$ to $d=171nm$. Because the difference in disorder between the samples is small and the optical quality is similar, this suggests that lasing is enhanced by the additional confinement of light provided by the PBG.  However, despite the reduction of the lasing threshold, the lasing output remains non-directional and similar to that of a random laser (figure \ref{fig:lasing} b). The absence of directionality of laser emission excludes the possibility of lasing in the defect states within the [111] PBG or band edge lasing in the high symmetry directions of the PhC.\cite{cao_lasing_2002,shkunov_tunable_2002,scharrer_ultraviolet_2006} Random lasing modes with enhanced confinement in the [111] direction would be expected to emit more strongly in other directions outside the PBG.  
However, the partial gap in the $171nm$ sample is located near the absorption edge and covers only a narrow solid angle relative to directions outside the gap, so the angular redistribution of light is small.  An alternative
explanation in terms of directional lasing plus subsequent diffusion of laser emission appears unlikely since a short absorption length of the pump light would make lasing occur close to the sample surface.  Based on these
experimental findings, we believe that what we observe is still random lasing and  the  spatial  confinement  of  the  random lasing modes is improved by reduced $k_c$ from the Bragg diffraction in the partial PBG.  Further
experimental and theoretical studies will be needed for a complete description of the complicated physical properties of such optically active, partially disordered photonic structures.  

\section{Conclusion}
   
We have demonstrated UV lasing at room temperature from ZnO inverse opal PhCs. The disorder in the structures, primarily due to imperfections in the opal templates and roughness caused by grain growth during firing, induces optical scattering and leads to random lasing.  When the first $\Gamma$L-pseudogap of the PhCs is tuned to the ZnO gain spectrum, a pronounced reduction in lasing threshold is observed, indicating enhanced confinement of light by the incomplete PBG.  We believe a fine tuning of the PBG position and a reduction in disorder by improving the sample fabrication will lead to photonic crystal lasing and a further reduction of lasing thresholds and improvement of the output directionality.  

\ack{This research project is supported by NSF Grant Nos. ECS-0823345, DMR-0808937, DMR-0814025 and the use of facilities of the NSF MRSEC program (DMR-0076097) at Northwestern University.}


\section*{References}

\begin{thebibliography}{2}

\bibitem{cao_lasing_2003}Cao H 2003 {\it Waves In Random Media} {\bf 13} R1--R39

\bibitem{noginov_solid_2005}Noginov M A 2005 {\it Solid-State Random Lasers} (Springer)

\bibitem{cao_review_2005}Cao H 2005 {\it J. Phys.} A {\bf 38} 10497--10535

\bibitem{wiersma_physics_2008}Wiersma D S 2008 {\it Nature Phys.} {\bf 4} 359--367

\bibitem{wu_random_2004}Wu X \etal 2004 {\it J. Opt. Soc. Am.} B {\bf 21} 159--167

\bibitem{Gottardo_2008}Gottardo S \etal 2008 {\it Nature Photon.} {\bf 2} 429--432

\bibitem{john_strong_1987}John S 1987 {\it Phys. Rev. Lett.} {\bf 58} 2486--2489

\bibitem{vlasov_different_1999}Vlasov Yu A \etal 1999 {\it Phys. Rev.} B {\bf 60} 1555--1562

\bibitem{hughes_2005}Hughes S \etal 2005 {\it Phys. Rev. Lett.} {\bf 94} 033903

\bibitem{koenderink_optical_2005}Koenderink A F \etal 2005 {\it Phys. Rev.} {\bf 72} 153102 

\bibitem{yamilov_highest-quality_2004}Yamilov A and Cao H 2004 {\it Phys. Rev.} A {\bf 69} 031803

\bibitem{yamilov_self-optimization_2006}Yamilov A \etal 2006 {\it Phys. Rev. Lett.} {\bf 96} 083905

\bibitem{rodriguez_disorder-immune_2005}Rodriguez A \etal 2005 {\it Opt. Lett.} {\bf 30} 3192--3194

\bibitem{kwan_transition_2007}Kwan K C \etal 2007 {\it Opt. Lett.} {\bf 32} 2720--2722

\bibitem{vlasov_enhancement_1997}Vlasov Yu A \etal 1997 {\it Appl. Phys. Lett.} {\bf 71} 1616--1618

\bibitem{maskaly_amplified_2006}Maskaly G R \etal 2006 {\it Adv. Mater.} {\bf 18} 343--347

\bibitem{romanov_stimulated_2004}Romanov S G \etal 2004 {\it Phys. Stat. Sol.} (c) {\bf 1} 1522--1530

\bibitem{yoshino_amplified_1999}Yoshino K \etal 1999 {\it Appl. Phys. Lett.} {\bf 74} 2590--2592
 
\bibitem{cao_lasing_2002}Cao W \etal 2002 {\it Nature Mater.} {\bf 1} 111--113

\bibitem{shkunov_tunable_2002}Shkunov M N \etal 2002 {\it Adv. Func. Mater.} {\bf 12} 21--26

\bibitem{scharrer_fabrication_2005}Scharrer M \etal 2005 {\it Appl. Phys. Lett.} {\bf 86} 151113

\bibitem{jurez_zno_2005}Jurez B H \etal 2005 {\it Adv. Mater.} {\bf 17} 2761--2765

\bibitem{scharrer_ultraviolet_2006}Scharrer M \etal 2006 {\it Appl. Phys. Lett.} {\bf 88} 201103

\bibitem{koenderink_experimental_2003}Koenderink A F \etal 2003 {\it Phys. Stat. Sol.} (a) {\bf 197} 648--661 

\bibitem{cao_random_1999}Cao H \etal 1999 {\it Phys. Rev. Lett.} {\bf 82} 2278--2281

%
%
%
%
%
%
%
%
%
%
%
%
%
%
%
%
%
%
%
%
%
%
%
%


\end{thebibliography}

\end{document}